\documentclass[11pt]{article} 
\usepackage{hyperref} 
\pdfoutput=1 
 
\usepackage{geometry}
\begin{document} 
 
\title{Wrinkly fingers: \\
the interaction between fluid- and solid-based \\ 
instabilities in elastic-walled Hele-Shaw cells
} 
 
\author{Draga Pihler-Puzovi\'{c}, Anne Juel and Matthias Heil\\ 
\\\vspace{6pt} Manchester Centre for Nonlinear Dynamics 
and School of Mathematics, \\ University of Manchester, M13 9PL Manchester, UK} 
 
\maketitle 
 
\begin{abstract} 
In this fluid dynamics video, we study a two-phase flow in 
an elastic Hele-Shaw cell that involves two distinct fluid- and solid-based instabilities:
viscous fingering and sheet buckling.
We show that the relative importance of the two instabilities
is controlled by a single non-dimensional parameter, which 
provides a measure of the elasticity of the flexible wall. We employ numerical
simulations to show that for relatively stiff [soft] walls, 
the system's behaviour is dominated by viscous fingering [sheet
buckling]. Strong interactions between the two instabilities arise in an intermediate
regime and lead to the development of extremely complex fingering and
buckling patterns.
\end{abstract} 
 
% main text 
 
Two identical videos were produced: A high quality \href{http://www.maths.manchester.ac.uk/~draga/DFDvideo2013/}{Video HQ}
and a lower quality video \href{http://www.maths.manchester.ac.uk/~draga/DFDvideo2013/}{Video LQ}.
 
We consider a circular Hele-Shaw cell of radius $R_{\rm {outer}}$, whose upper boundary is formed
by an elastic membrane of thickness $h$, Young's modulus $E$ and Poisson's ratio $\nu$.
In the initial state, the cell has uniform depth $b_0 \ll R_{\rm {outer}}$, and is filled with a viscous fluid 
of dynamic viscosity $\mu$ and surface tension $\gamma$.
We then inject air at a constant flow rate $\dot{V}$ at the centre of the cell. 
The air displaces the viscous fluid and inflates the elastic sheet$^{[2], [3], [4]}$.
 
We study the behaviour of the system using the depth-averaged lubrication equations for the fluid flow, 
coupled to the F\"{o}ppl-von K\'{a}rm\'{a}n equations which describe the deformation of the thin elastic wall; 
these equations describe both stretching and bending of the bounding elastic sheet. 
We also account for the pressure jump across the curved air-liquid interface induced by surface-tension.
The problem
is governed by four non-dimensional parameters: the aspect ratio of the cell
$\mathcal{A} = b_0/R_{\rm {outer}}$; the capillary number $\mathrm{Ca} = \mu \dot{V}/2\pi b_0 \gamma R_{\rm {outer}}$; 
the F\"{o}ppl-von K\'{a}rm\'{a}n parameter $\eta = 12(1-\nu^2)(R_{\rm {outer}}/h)^2$, which 
controls the relative importance of in-plane and bending stresses during the elastic sheet deformation; 
and the fluid-structure interaction (FSI) parameter $\mathcal{I} = 144 (1-\nu^2) \mu \dot{V} / 2\pi \mathcal{A}^3 E h^3$, 
which represents the ratio of
the typical viscous stresses in the fluid to the bending stiffness of the elastic membrane. 

In the fluid dynamics video, we fixed the aspect ratio and the capillary number to $\mathcal{A} = 0.04$
and $Ca = 0.4$, respectively, and set the F\"{o}ppl-von K\'{a}rm\'{a}n parameter to
$\eta = 2 \times 10^6$.
We varied the FSI parameter $\mathcal{I}$ (which corresponds to changing the Young's modulus of
the sheet while keeping all other parameters constant) to assess the effect of fluid-structure
interaction on the fingering and buckling instabilities.
We explored interactions between these instabilities by direct numerical simulation, using a
finite element discretization of the governing equations implemented in the open-source 
finite element library \texttt{oomph-lib}$^{[1], [3], [4]}$. 
The initial configuration for all simulations is zero wall displacement and a circular bubble with 
radius $R_{\rm init} = 0.05 R_{\rm outer}$.
We perturbed the bubble radius by a small non-axisymmetric
perturbation with amplitude of 1\% of the initial radius and wavenumber 7 (found to be 
the most unstable wavenumber in the corresponding rigid-walled cell).

The values of the parameter $\mathcal{I}$ in the video are $10^3$, $10^4$, $10^5$ and $10^6$. 
An increase in $\mathcal{I}$  increases the transverse deflection of the membrane, and weakens (and
ultimately suppresses) the viscous fingering instability. 
However, for sufficiently thin-walled
bounding membranes an increase in $\mathcal{I}$ also increases their propensity to wrinkle, resulting
in intriguing competition (and interaction) between two distinct symmetry-breaking
instabilities: for small values of $\mathcal{I}$ the system’s behaviour is dominated by viscous fingering,
with the non-axisymmetric deformation of the bounding membrane arising as a passive
response to the spatial variations in the fluid pressure. In an intermediate range of $\mathcal{I}$, the
(weakened) fingering instability initially develops independently of the wrinkling process
but the two instabilities interact strongly when their amplitudes reach finite values. Finally,
for large values of $\mathcal{I}$ , viscous fingering is completely suppressed and the non-axisymmetric
deformation of the expanding air-liquid interface arises as a passive response to the wrinkling
of the membrane, reversing the role of the two instabilities. 

With this video, we demonstrate the complexity of the interaction between the two distinct instabilities,
viscous fingering and sheet buckling, by varying a single non-dimensional parameter. Qualitatively similar
behaviour was observed in the laboratory experiments with polypropylene sheets ($h = 30 \mu$m, $E=3.6$ GPa)
in the cell with $b_0 = 0.57$ mm and $R_{\rm outer} = 17$ cm, filled with 1000 cSt silicone oil. \\ \\

\noindent\textbf{References}\\

\noindent [1] M. Heil and A. L. Hazel. $\mathtt{oomph-lib}$ - an object-oriented multi-physics finite-element
library. In M. Sch\"{a}fer and H.-J. Bungartz, editors, \textit{Fluid-Structure Interaction}, pages 19-49. 
Springer, 2006. \\
$\mathtt{oomph-lib}$ is available as open-source software at $\mathtt{http://www.oomph-lib.org}$.\\

\noindent [2] D. Pihler-Puzovi\'{c}, P. Illien, M. Heil, and A. Juel. Suppression of complex fingerlike
patterns at the interface between air and a viscous fluid by elastic membranes. \textit{Phys.
Rev. Lett.}, 108:074502, 2012.\\

\noindent [3] D. Pihler-Puzovi\'{c}, A. Juel, and M. Heil. The interaction between viscous fingering and
wrinkling in elastic-walled Hele-Shaw cells. \textit{submitted to Phys. Fluids}, 2013.\\

\noindent [4] D. Pihler-Puzovi\'{c}, R. P\'{e}rillat, M. Russell, A. Juel, and M. Heil. Modelling the suppression
of viscous fingering in elastic-walled Hele-Shaw cells. \textit{J. Fluid Mech.}, 731:162-183,
2013.

\end{document}